\documentclass[conference]{IEEEtran}
\IEEEoverridecommandlockouts
\pdfoutput=1
\usepackage{cite}
\usepackage[utf8]{inputenc}
\usepackage{algorithm}
\usepackage{amsmath,amssymb,amsfonts}
\usepackage{subfig}
\usepackage{graphicx}
\usepackage{textcomp}
\usepackage{amsmath}
\usepackage[outdir=./]{epstopdf}

\usepackage[noend]{algpseudocode}
\usepackage{xcolor}
\def\BibTeX{{\rm B\kern-.05em{\sc i\kern-.025em b}\kern-.08em
    T\kern-.1667em\lower.7ex\hbox{E}\kern-.125emX}}
\begin{document}

\title{Parallelization of Kmeans++ using CUDA\\
}

\author{\IEEEauthorblockN{1\textsuperscript{st} Maliheh Heydarpour Shahrezaei}
\IEEEauthorblockA{
\textit{Pouyandegane-danesh 
higher education institute}\\
Chalus, Iran \\
maliheh.heydarpour@pd.ac.ir}
\and
\IEEEauthorblockN{2\textsuperscript{nd}  Reza Tavoli}
\IEEEauthorblockA{
\textit{Islamic Azad University}\\
Chalus, Iran \\
R.tavoli@iauc.ac.ir}}

\maketitle

\begin{abstract}

K-means++ is an algorithm which is invented to improve the process of finding initial seeds in
K-means algorithm. In this algorithm, initial seeds are chosen consecutively by a probability which
is proportional to the distance to the nearest center. The most crucial problem of this
algorithm is that when running in serial mode, it decreases the speed of clustering. In this paper,
we aim to parallelize the most time consuming steps of the k-means++ algorithm. Our purpose
is to reduce the running time while maintaining the quality of the serial algorithm.
\end{abstract}

\begin{IEEEkeywords}

K-means++, Parallelization, Clustering
\end{IEEEkeywords}

\section{Introduction}

In the K-means clustering algorithm \cite{1}, a well-known version of clustering proposed by Lloyd \cite{13}, one needs to first determine the initial set for
K seeds. Generally, this is done by a random function, and the seeds are selected randomly. 
Many approaches are invented to ameliorate the process of finding the
initial seeds in the K-means algorithm. K-means++ \cite{2} algorithm is one of the methods that can
be used to find the centroids.

The K-means++ method can ameliorate the process of finding
initial seeds in K-means and can improve the quality of clusters by providing a more efficient
approach than the random method which is used in the K-means algorithm. In the K-means++ approach,
the seeds are chosen based on a probabilistic method where any point x is selected with a
probability which is proportional to the square of the distance of x.

Because of the importance of clustering algorithms  in different areas of studies  \cite{10, 17, 18, 21} and can be used for designing various intelligent systems \cite{11}. They can even be utilized to provide anonymity for the networks to further improve the methods like \cite{19, 20}. Considering the preceding fact,  devising a methodology for accelerating the clustering process is significant. Several methods have been invented to parallelize the clustering algorithms \cite {14, 16, 22} including k-means \cite{7, 8, 9} to further improve the efficiency of them. However, they are a few works on parallelization of K-means++. 

The earliest approach is a master’s thesis by Karch which was revealed in 2010
\cite{3}. The thesis talks about parallelization of both K-means and K-means++ over GPU. However, this only considers the parallelization of the initial part of the algorithm and none
of its other parts. Bahmani et al. also discuss the K-means++ method in \cite{4}. They present
a different K-means initialization technique, however, they do not provide
time comparisons between their method and K-means or even K-means++.

In the presented work, we aim
to provide the implementation for parallelization of K-means++. Our implementation is over GPU
 and we use different approaches for decreasing the time of processing. Finally, we present our
results to illustrate how parallelization can significantly improve running time in comparison to
serial mode.


\section{K-means++ Parallelization}
\label{sec2}
  A pseudo-code of serial implementation of K-means++ has been shown in Algorithm \ref{euclid}. Algorithm \ref{euclid} shows that, after the selection of each seed, the distance of each point to the nearest seed is updated.

\begin{algorithm}[h]
		\caption{Serial K-means++}
		\label{euclid}
		\begin{algorithmic}[1]
			\Procedure{MySerialK-Means++}{}
			\State \textbf{Input}:\\
			~~~~~~~~H: List of Data points(N), Number of
			
			   ~~Clusters(K), Distance Function (d)
			\State \textbf{Output}:\\  
             ~~~~~~~~O: A set of points n, classifies into

             ~~k different clusters
			\State \textbf{Function Serial-K-means++(N, K):}\\
             ~~~ \textit{//select the first seed randomly}\\
             ~~~$H_m(O) = rand~~ n \in N$\\
             ~~~$O(l) = \infty$\\
             ~~~\textit{ // Rest of the seeds selected by distance and \\~~~by probability}\\
             
             ~~~Define \textit{m, l:}\\
			
			\For { $m$ $\in$ $(1, k-1)$}
			\For {$l$ $\in$ $(1, n)$}\\
			 ~~~~~~~~~~~//\textit{Update the nearest distance}\\
              ~~~~~~~~~~$d_{m, l}= d(H_m n_l)$\\
              ~~~~~~~~~~$d_{sn}= [[min_{1, x, n}](d_{x, l})]^2$\\
              ~~~~~~~~~//Taking $H_{m+1}$\\
              ~~~~~~~~~//\textit{Function Probability of n ($d_{sn}$, $n_l$)}\\
              ~~~~~~~~~$P(n_l) = \frac{d_{sn}}{\sum_{w=1}^{w =z} d_{sn_{w}}}$\\
			\EndFor
			\EndFor
			\State Return \textbf{$H$}
			\EndProcedure
		\end{algorithmic}
	\end{algorithm}

 In the following, we describe the method for parallelization of K-means++. As we previously
mentioned, we aim to parallelize the most time consuming steps of K-means++ algorithm. As it
is indicated in algorithm \ref{euclid}, in K-means++ algorithm distance of all points to the chosen center
should be calculated and then the seeds are chosen based on a probability formula. These are
the most challenging aspects for parallelizing the K-means++. In Algorithm \ref{euclid1}, we have shown
the process of parallelization of the K-means++ algorithm. As Algorithm \ref{euclid1} illustrates,
the parallelization procedure starts with partitioning the points into $l$ different points. It then
continues to parallelize the calculation of the distance of points to the centers until the end of
the algorithm.

\begin{algorithm}[h]
		\caption{Parallel K-means++}
		\label{euclid1}
		\begin{algorithmic}[1]
			\Procedure{MyParallelK-Means++}{}
			\State \textbf{Input}:\\
			~~~~~~~~H: List of Data points(N), Number of
			
			   ~~Clusters(K), Distance Function (d)
			\State \textbf{Output}:\\  
             ~~~~~~~~O: A set of points n, classifies into

             ~~k different clusters
			\State \textbf{Function Parallel-K-means++(N, K):}\\
             ~~~ \textit{//select the first seed randomly}\\
             ~~~$H_m(O) = rand~~ n \in N$\\
             ~~~$O(l) = \infty$\\
             ~~~\textit{ // Rest of the seeds selected by distance and \\~~~by probability}\\
             
             ~~~Define \textit{m, l:}\\
		    ~~~Partition N ($N_1, \dots, N_l$)\\
			\For { $m$ $\in$ $(1, k-1)$}
			\For {$l$ $\in$ $(1, n)$} $\rightarrow$ \textbf{In Parallel}\\
			 ~~~~~~~~~~~//\textit{Update the nearest distance}\\
              ~~~~~~~~~~$d_{m, l}= d(H_m n_l)$\\
              ~~~~~~~~~~$d_{sn}= [[min_{1, x, n}](d_{x, l})]^2$\\
              ~~~~~~~~~//Taking $H_{m+1}$\\
              ~~~~~~~~~//\textit{Function Probability of n ($d_{sn}$, $n_l$)}\\
              ~~~~~~~~~$P(n_l) = \frac{d_{sn}}{\sum_{w=1}^{w =z} d_{sn_{w}}}$\\
			\EndFor
			\EndFor
			\State Return \textbf{$H$}
			\EndProcedure
		\end{algorithmic}
	\end{algorithm}

In order to start the parallelization process,we get help from CUDA \cite{12}, which is a parallel computing platform designed by Nvidia \cite{23}. CUDA is used to accelerate different applications in various areas like computational chemistry, bioinformatics  and other fields of research \cite{24, 25, 26, 27, 28}.

 Using CUDA, we consider several different methodologies,  for the parallelization including:

\begin{itemize}
  \item Global Memory
  \item Constant Memory
  \item Texture Memory
  \item Parallel reduction using Thrust
\end{itemize}

We should mention that we also consider using shared memory, kernel launch overhead and loop
unrolling methods in our proposed method for applying further methods of parallelization, but these three approaches had some problems to be applied on our work. On the one hand, when using shared memory, all of the blocks need to fetch a common piece of data from shared memory. In this work, the only parameters that could be applied over shared memory are centroids. However, fetching centroids one by one from shared memory
for several blocks would not be efficient, and some alternatives like Constant memory would
be more helpful. On the other hand, loop unrolling was not applicable, since the parameters in the for  loop were changeable. So it was not possible to apply this
method over the algorithm. Moreover, in the case of kernel launch overhead we analyze and see that
the amount of processing we are doing in each kernel launch is much more than the overhead of
launching kernel.  This is the reason the kernel launch overhead is insignificant as well.

As a result, we consider using Global memory, Constant memory, Texture memory, and a parallel
reduction in thrust \cite{5} for doing parallelization. Furthermore, the focus of our work is
on the initial centroid selection. The clustering part is implemented the same as the K-means algorithm.
In fact, in parallel mode, the initialization part takes more time to be executed rather than the
clustering part. In the centroid selection process, every time a new centroid is found, it’s index
is stored in a separate array. We use these indices to find the coordinates of the centroid in
the array containing all the points, and in case of shared memory experiment, its coordinates are
stored in Constant memory for fast retrieval. 

Each thread in the centroid selection process is
responsible for calculating the distance of a single point from all the selected centroids; hence,
each thread has an equal workload. This workload increases equally with each subsequent centroid
selection. If the K is small, the workload of each thread is small and equal at all times.
Starting the parallelization, first, we used Global memory. We allocate all of the data points to
reside on Global memory. Afterward, we define the block of threads, and each thread is used for
calculation of distances to a chosen centroid. In our implementation, we consider the following
configuration for the size of blocks and the number of blocks per thread:

\begin{equation}
Thread = 1024
Block = ceil( \frac{number of points}{threads})
\end{equation}

Using Global memory, we could improve the performance of the algorithm by a large margin.
Next to using Global memory, we use Constant memory. Constant memory can be used for data
that will not change when the kernel is executing. Constant memory is a read-only memory, and it is
faster than Global memory. Using Constant memory in place of Global memory, we can reduce the
memory bandwidth in some cases. NVIDIA hardware has provided 64KB Constant memory which
its space is cached. The advantage of this kind of memory is that, when threads of half warp read
from constant cache, it would be as fast as reading from the register if all of the threads are reading
the same address. Considering the preceding facts, we try to use Constant memory as a method
for amending the efficiency of K-means++ algorithm. 

Using Constant memory, we first decide to
put all of the points to reside on Constant memory. However, when the number of points
increases by a large margin, the algorithm does not work. This is because there are too many points for the amount of Constant memory. As we previously discussed,
Constant memory is only 64kb while we are going to consider more than millions of points when we
are evaluating our method. As a result, we should put only the centroids in Constant memory, instead of putting the points. Using Constant memory will better improve the performance of the
algorithm. If we increase the num clusters from 30 to 50, there is a slight improvement in Constant
memory version. If we further increase from 50, it illustrates much more improvement. We will
show these results in section Sec~\ref{sec3}.

Next to Constant memory and Global memory, we further expand the methodology for parallelization of the K-means++ algorithm by using Texture memory. This type of memory is as like as
Constant memory. It can provide better efficiency and reduce memory traffic if accessing and
reading from memory follow a specific pattern. It is also similar to Global memory, but
it performs better because it is treated as read-only memory. Since the Texture memory cannot
be modified from the device side, the device does not need to keep track of updates and cache
coherence. We should mention that generally reading from Global memory is faster than Texture
memory if the access has coalesced. However, if we are accessing to the neighbors in memory locations, using Texture memory will give us better results than Global memory. In fact, in specific
addressing locations, reading from device memory using Texture memory and fetching from that,
would be more beneficial than fetching from Global or even Constant memory. 

Since we are
accessing the same cells in memory, we will use Texture memory, which can perform more efficiently than Global memory and can improve the performance of the K-means++ algorithm. The approach here is to store all the points in Texture memory
rather than Global memory because their location does not change.
Apart from using Global, Constant, and Texture memory, we use parallel reduction method.
Since we have Sum() function in the K-means++ algorithm, we can rely on using the partial sum
method for improving the performance. In order to do this, we use Thrust. Thrust is a library
based on C++ for CUDA which is based on Standard Template Library (STL). It provides two
different variables call host\underline{\space}vector and device\underline{\space}vector. As the name shows, host\underline{\space}vector has resided in host
memory, but the device\underline{\space}vector is stored in device memory. They both can be added as headers in the
source code as :

$\#$include $<$ thrust/host$\_$vector.h $>$\\$~~~$
$\#$include $<$ thrust/device$\_$vector.h $>$

In order to make the reduction, a reduction algorithm utilizes binary operation, and further, it tries to reduce
any specific input sequence to a single value. For instance, when summing sequences of numbers,
one can reduce the arrays for each sum operations. In Thrust, partial sum or sum of arrays in
reduction mode can be implemented by using Thrust reduce function (thrust::reduce). Following
is the way that Thrust implements reduce for Sum() function:

intsum = thrust :: reduce(D.begin(), D.end(),(int)0, thrust :: $~~~$ plus $<$ int $>$ ());

Above, D.begin(), D.end() identify the range for values and (int)0, thrust :: plus $<$ int $>$ ())
describe the initial value and the reduction operator, which is plus in this case. Considering this,
in our implementation, we use the following arguments for making sum reduction:

doublesum = reduce(dptr, dptr + num points, 0.0);

We should also mention that using Thrust for reduction is common
in all of the experiments and is assumed to perform continuously. Therefore, the performance
of the overall algorithm for each experiment is not impacted by a reduction step differently, and the
overall performance can be considered as a measure of centroid selection. In the section Sec~\ref{sec3}, we
will talk about the performance improvement of K-means++ algorithm using the abovementioned methodologies.

\section{Results}
\label{sec3}
In this section, we will explain the evaluation part and analysis of the performance of the Kmeans++ algorithm in both parallel mode (GPU) and serial mode (CPU). Our implementation
is done over amazon web service (AWS) on Linux Ubuntu 16.04 and NVIDIA 367.27 driver.

\begin{figure}[h]\centering
\subfloat[Increasing number of clusters]{\label{1a}\includegraphics[width=.50\linewidth]{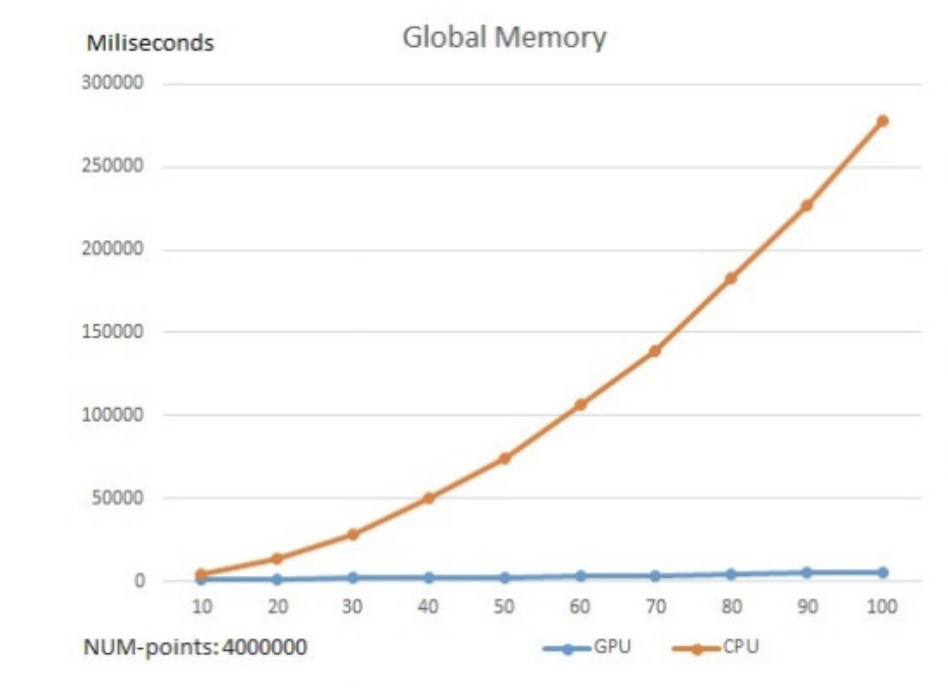}}\hfill
\subfloat[Increasing number of points]{\label{1b}\includegraphics[width=.50\linewidth]{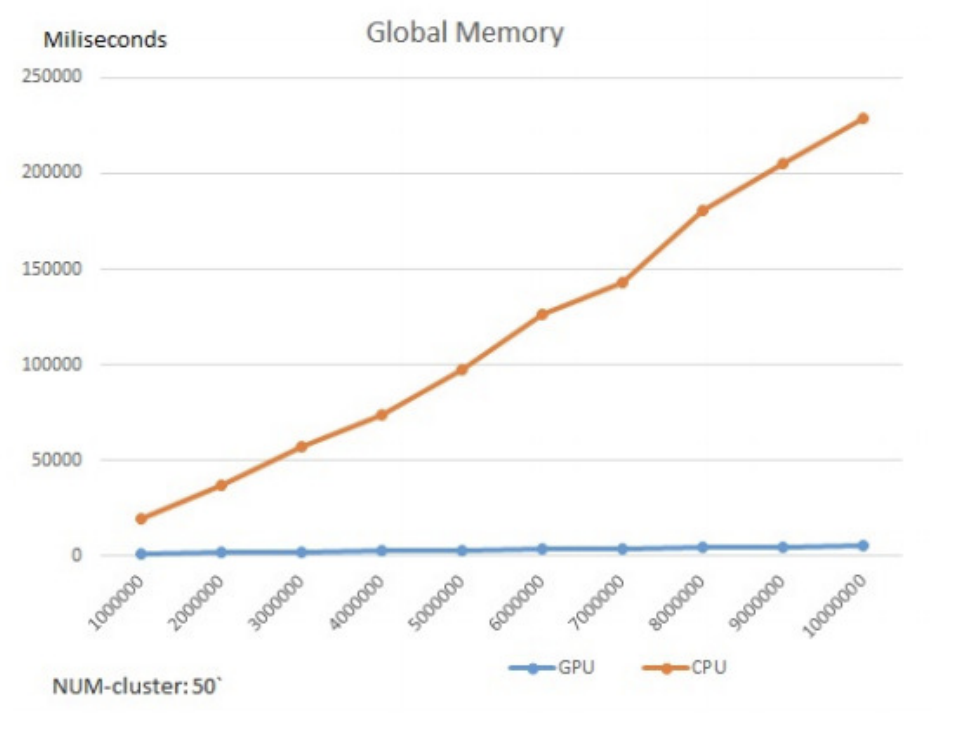}}\par 
\caption{Improvement over performance using Global Memory}
\label{fig1}
\end{figure}


We demonstrate the results for applying Global memory in our implementation. Using Global memory, we consider two scenarios: First, we discuss the improvements over GPU using Global memory rather than CPU when the number of points is increasing from 1 million to 10 million points, and the number of clusters is equivalent to a constant number (NumC = 50). Second, we explain the scenario when the number of clusters is increasing from 10 to 100 clusters, while the number of points is equal to a constant number (NumP = 4000000). In Fig.~\ref{1a} and Fig.~\ref{1b}, the results of using Global memory in abovementioned scenarios are indicated.
As it is shown in Fig.~\ref{fig1}, the performance of K-means++ algorithm has been amended when it is implemented over GPU than CPU. We can see more improvement when the number of points and clusters are increased gradually. Using Global memory, when the number of clusters grows from 10 to 100, GPU time varies between 1-5 while CPU time increases from 3 seconds to 4 minutes.

\begin{figure}\centering
\subfloat[Increasing number of clusters]{\label{2a}\includegraphics[width=.50\linewidth]{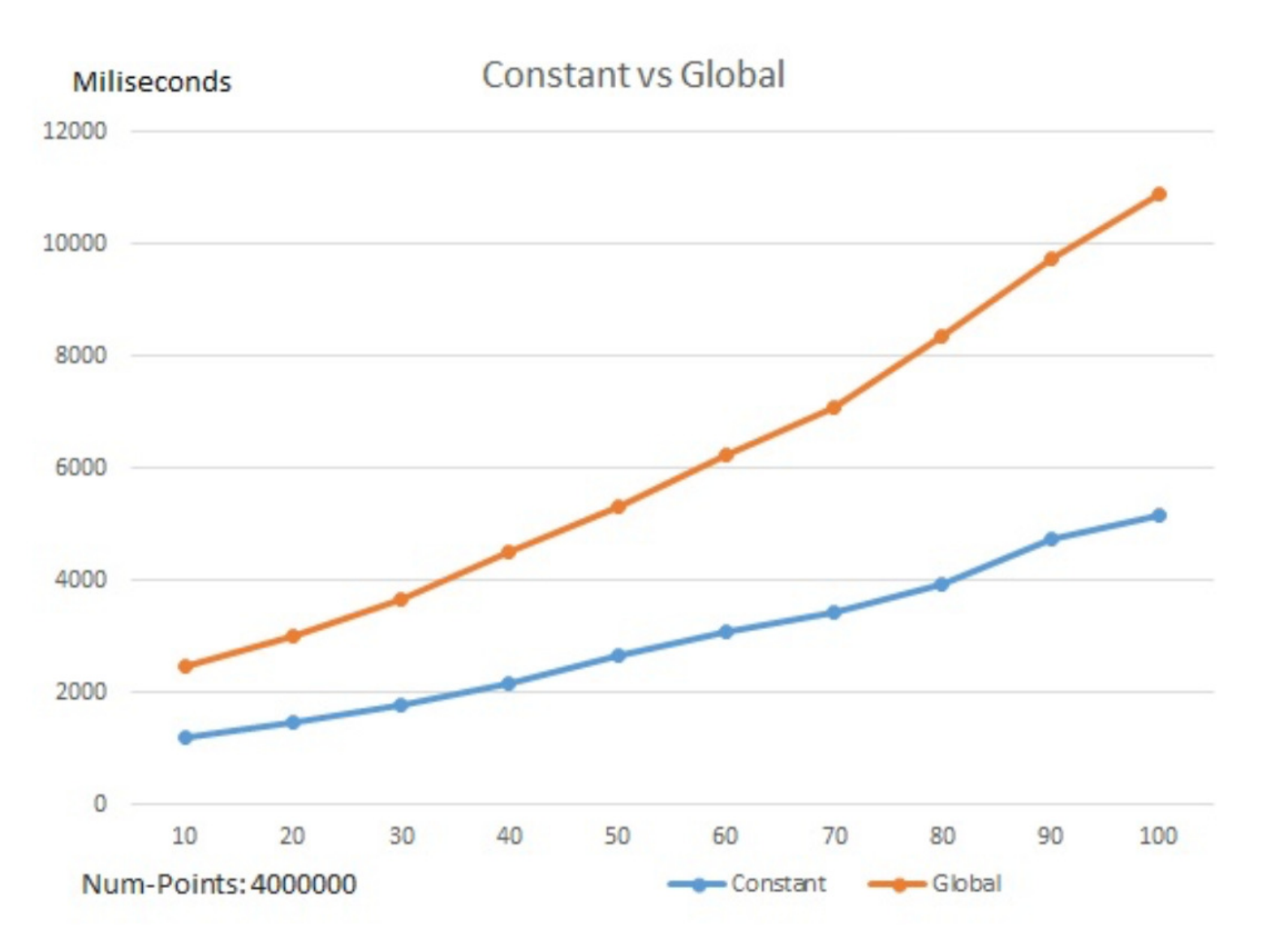}}\hfill
\subfloat[Increasing number of points]{\label{2b}\includegraphics[width=.50\linewidth]{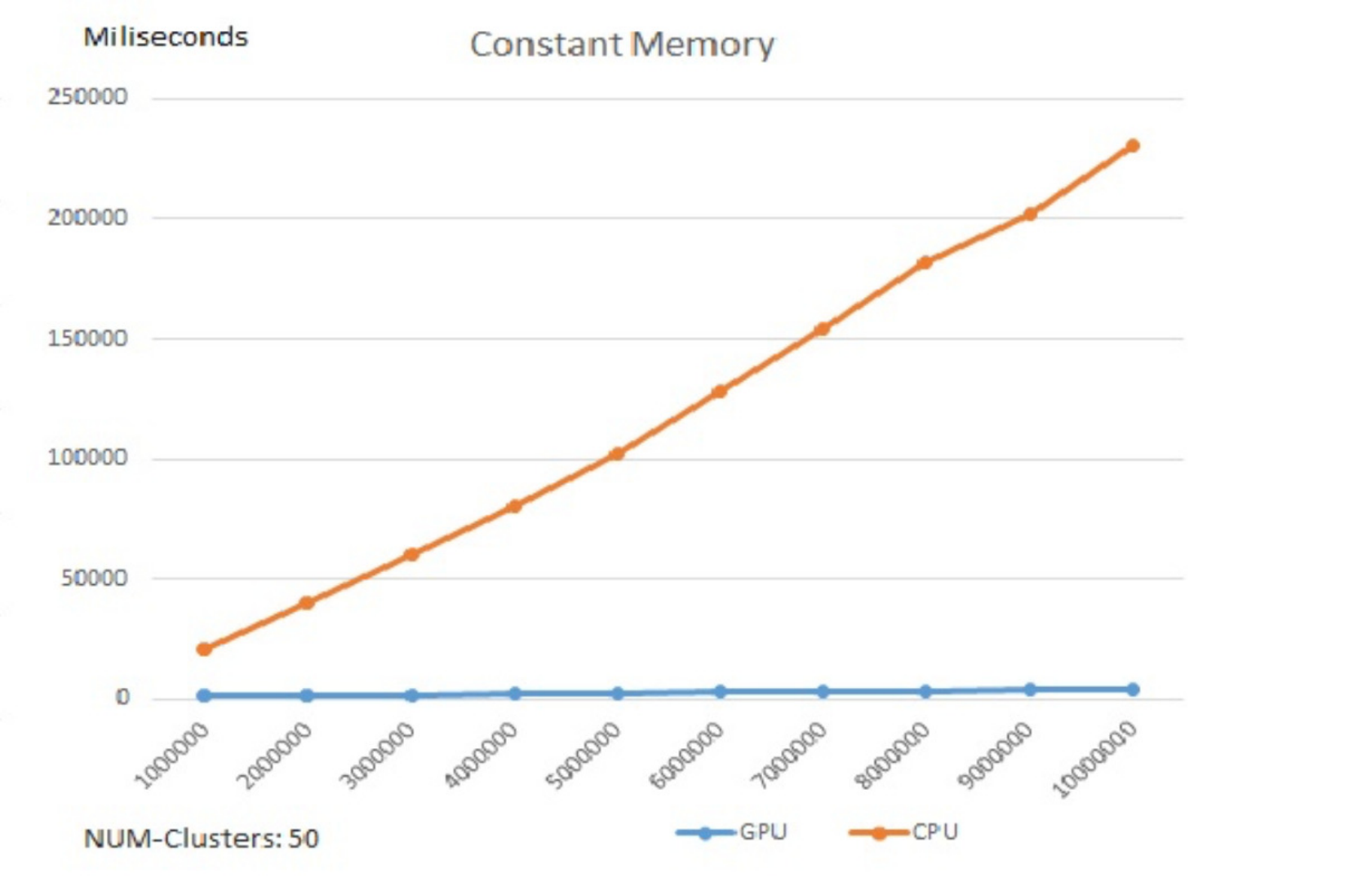}}\par 
\subfloat[Global vs Constant Memory]{\label{2c}\includegraphics[width=.50\linewidth]{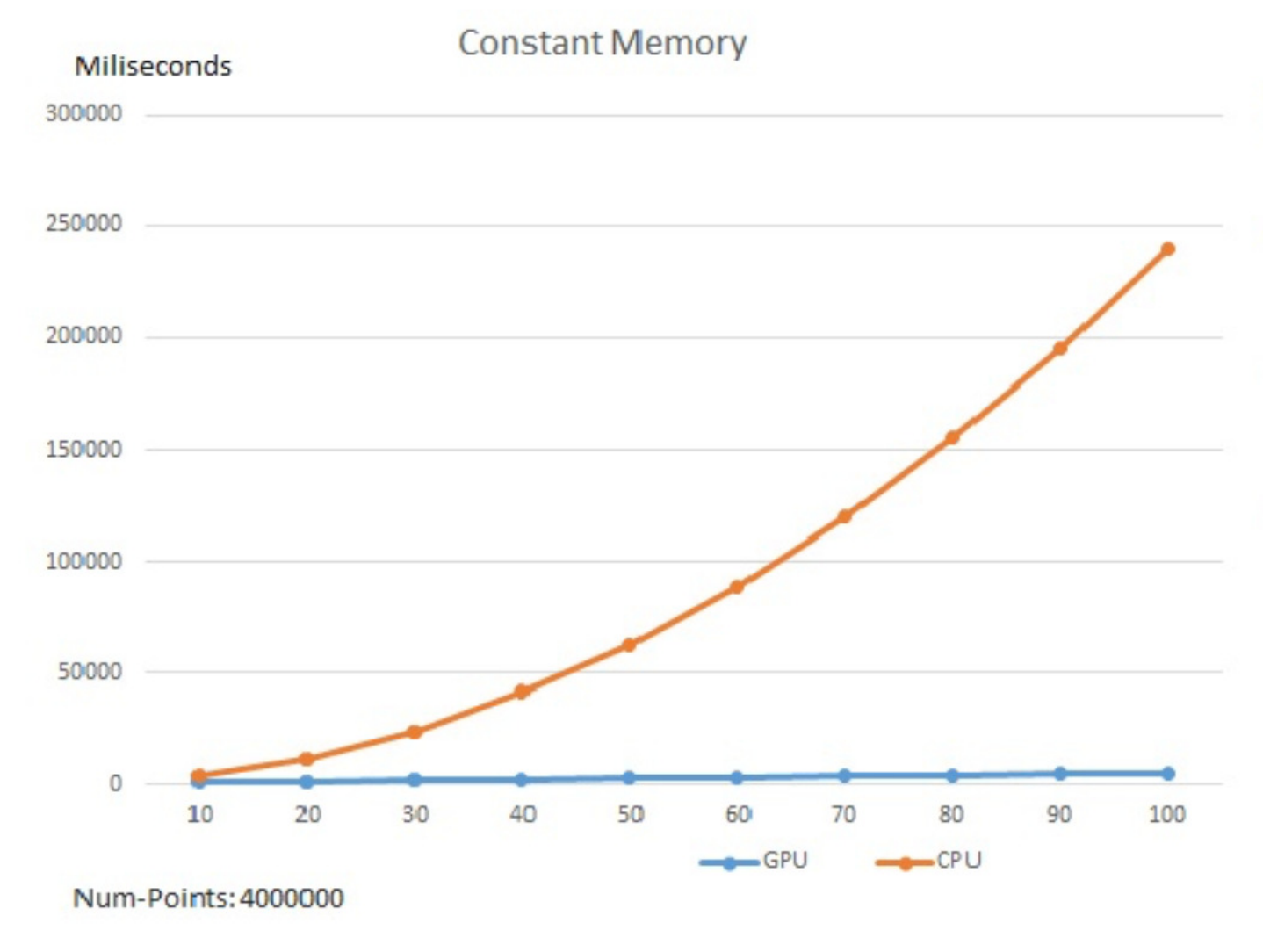}}
\caption{Improvement over performance using Constant Memory}
\label{fig2}
\end{figure}

On the other hand, increasing the number of points from 1-10 million, the GPU time changes within a range from 1-5 seconds while this time for the CPU version is from 18 seconds to 3 minutes.
Next to Global memory,  Fig.~\ref{2a} and Fig.~\ref{2b}, demonstrate the performance of the algorithm in parallel mode when we use Constant memory. As it is shown in Fig.~\ref{fig2}, using Constant memory, we have a piecemeal improvement in the performance of the algorithm. Increasing the number of clusters from 1 to 100, the GPU time changes within a range between 1-5 seconds while this time for CPU is between 3 seconds to 4 minutes. In case of growth from 1 to 10 million points, the GPU time is 1-5 seconds while the CPU time is 18 seconds to 3 minutes. One point that should be mentioned is that the performance of Constant memory and Global memory seems to be the same when we describe the time in seconds, however, this changes when we consider time in milliseconds. To better show this, in Fig. ~\ref{2c} we make a comparison between the performance of Constant memory and Global memory. As it is illustrated in Fig.~\ref{2c}, using Constant memory can provide better performance for the K-means++ algorithm. Comparing Global to Constant memory and changing the clusters from 10-100 points, we have slight improvement from 2-11 percent during these steps. We can even see better improvements when the number of clusters is further increased.

\begin{figure}\centering
\subfloat[Increasing number of clusters]{\label{3a}\includegraphics[width=.50\linewidth]{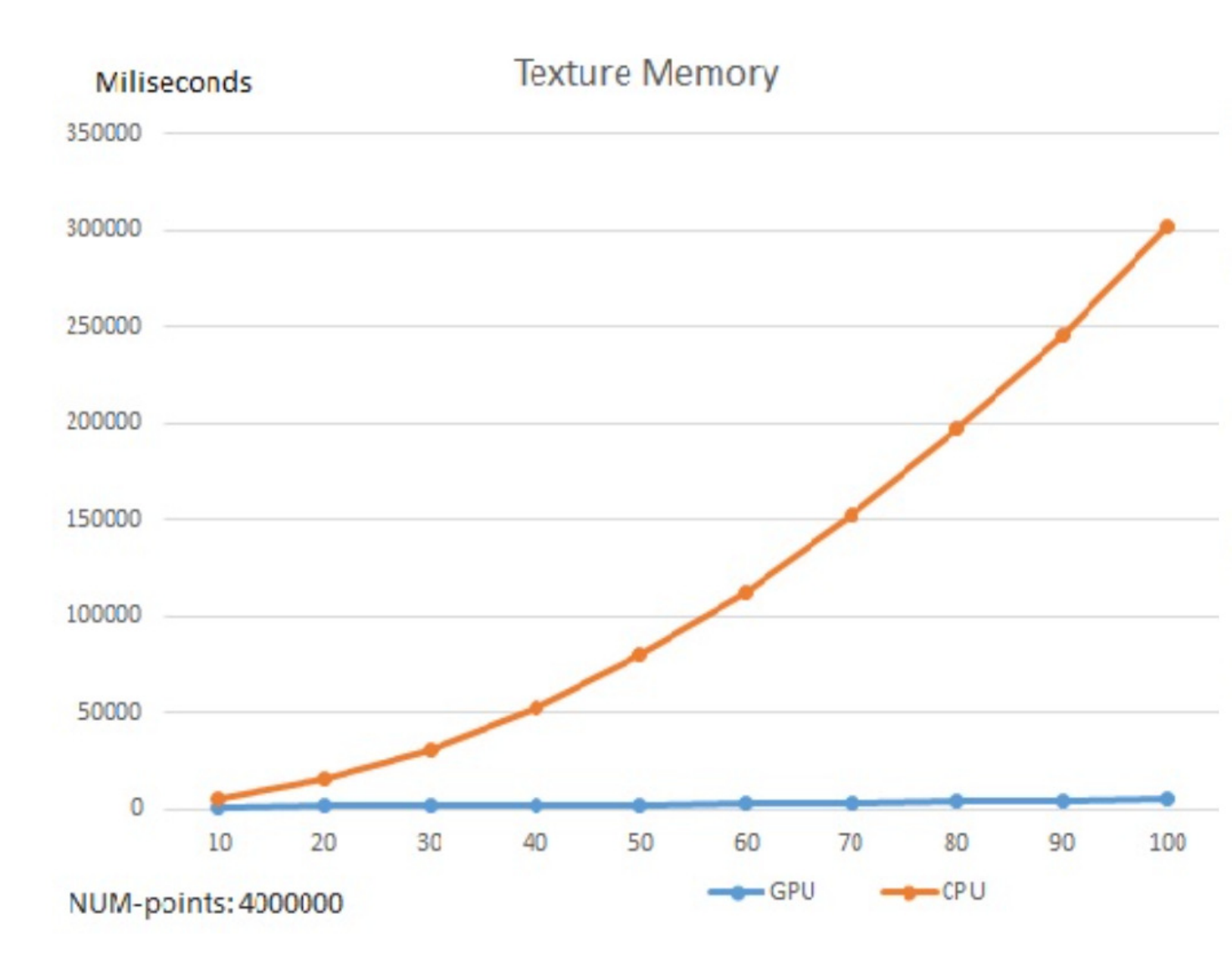}}\hfill
\subfloat[Increasing number of points]{\label{3b}\includegraphics[width=.50\linewidth]{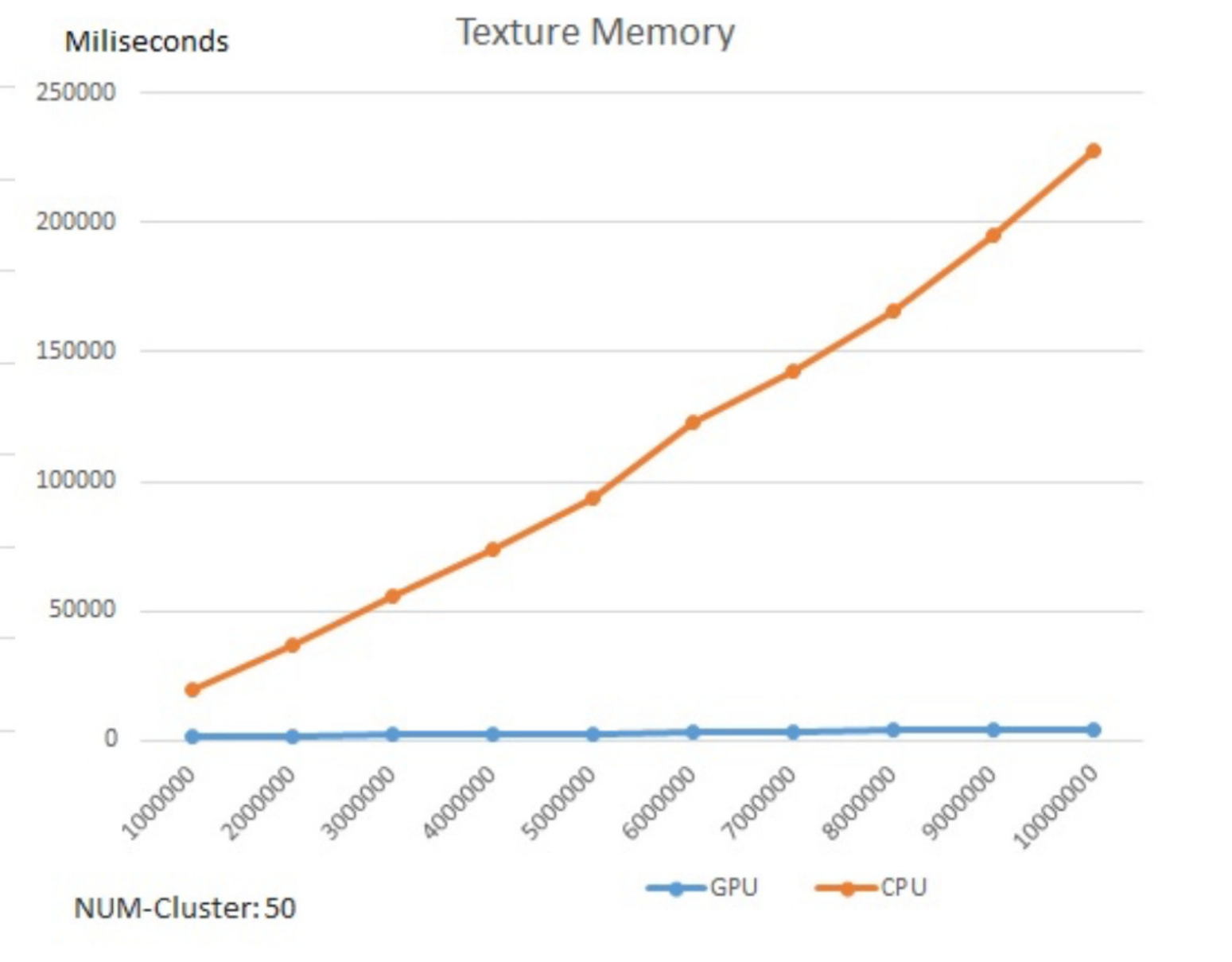}}\par 
\subfloat[Global vs Texture Memory]{\label{3c}\includegraphics[width=.50\linewidth]{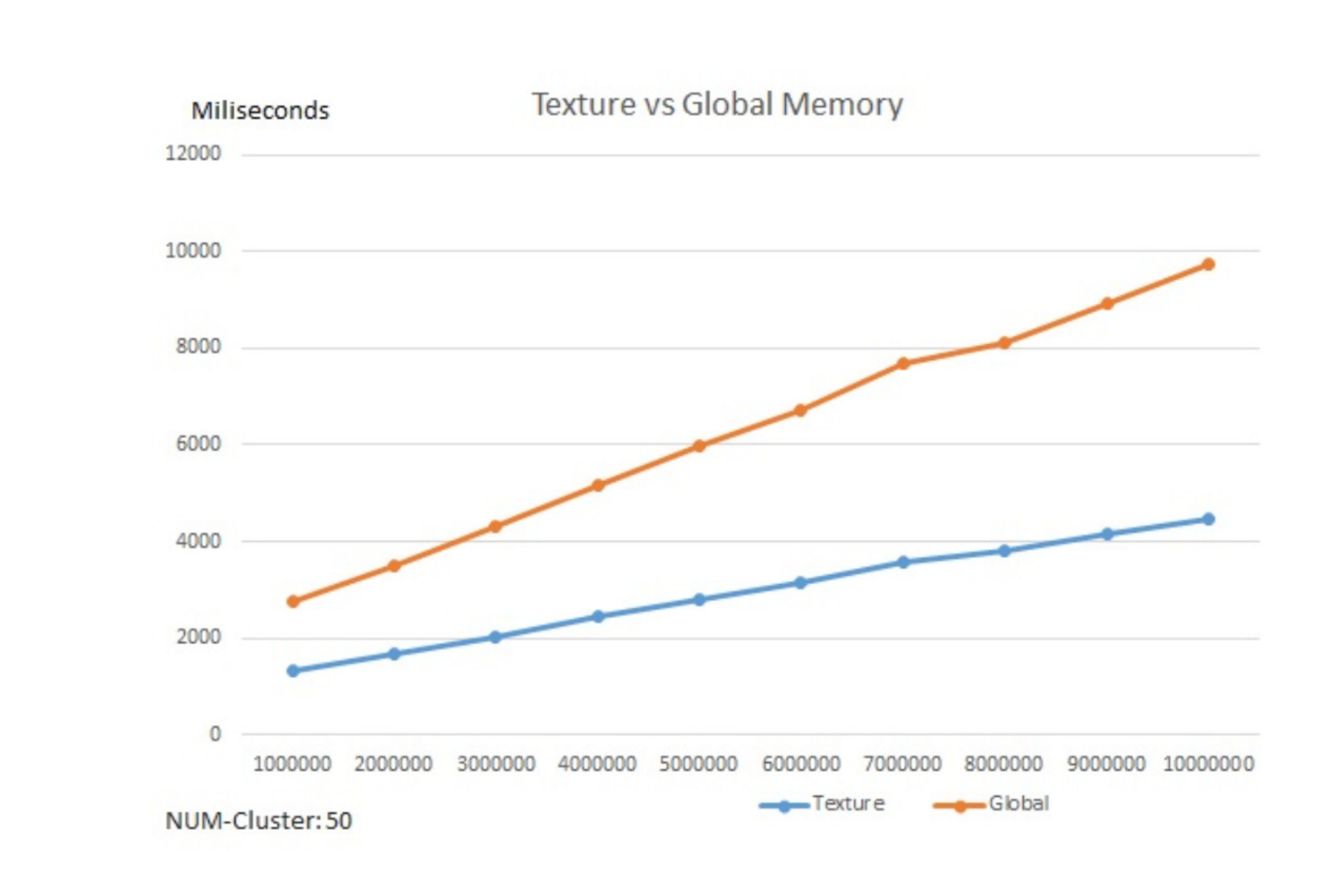}}
\caption{Improvement over performance using Texture Memory
}
\label{fig3}
\end{figure}

Beside Global and Constant memory, we evaluate the proficiency of the K-means++ algorithm using Texture memory. We apply both of the aforementioned scenarios for this type of memory. Results are shown in Fig.~\ref{fig3}. As it is indicated, using Texture memory, we have outstanding improvement and efficiency for the K-means++ algorithm. Incrementing number of clusters from 1 to 100 (Fig.~\ref{3a}), the GPU time is within range roughly similar to Global memory and CPU time changes from 3 seconds to 4 minutes. In case of increasing the number of points from 1 million to 10 million (Fig.~\ref{3b}), the GPU time changes between 1-4 seconds, which is slightly better than Global memory and the CPU time is the same as both Global and Constant memory. As we previously discussed, Texture memory can perform better when we are accessing neighbor locations in memory and because it is not modified from the device side. In  Fig. ~\ref{3c} we provide the time comparison between Global memory and Texture memory. As it is presented, each time Texture memory is performing better, and at each step, Texture memory is between 10 to 14 percent more efficient in performance than global memory.

One result is expected. When the number of clusters is smaller, (K $<$ 7) the performance of
CPU is slightly better than the GPU. This implies two things. First The data transfer from host to device and kernel launch overhead causes the GPU version takes longer. Second,  we work in a two-dimensional environment. If the dimension is larger, we can easier reach to the point that GPU can provide better speed. As we can see, there is a trade-off between the speed of the algorithm and the required workload.

\section{Conclusion}
\label{seccon}
In this work, we provide a comprehensive study over parallelization of K-means++ algorithm. First, we describe the methodologies of parallelization, which were feasible for being implemented over the algorithm and talk about the ones which were impossible to be implemented. Second, we provide a detailed explanation over implementing each of these methodologies and the way we used in our methods. Finally, we evaluate our work when using any of these methodologies and make a comparison between them. Our results show the efficient performance of the algorithm over GPU rather than CPU. Also, it was demonstrated that in the case of our work, Constant memory and Texture memory had better efficacy than Global memory.

\end{document}